\documentclass[prl,showpacs,preprintnumbers,amsmath,amssymb,twocolumn]{revtex4}

\usepackage{graphicx}
\usepackage{dcolumn}
\usepackage{bm}
\usepackage{color}
\usepackage{layout}
\usepackage{verbatim}

\newcommand{\ket}[1]{\left|#1\right>}
\newcommand{\bra}[1]{\left< #1 \right|}
\newcommand{\beq}{\begin{equation}}
\newcommand{\eeq}{\end{equation}}
\newcommand{\bep}{\begin{pmatrix}}
\newcommand{\eep}{\end{pmatrix}}

\newcommand{\Vr}{V_{r}}
\newcommand{\Vm}{V_{m}}
\newcommand{\Vl}{V_{l}}
\newcommand{\JL}{J_{l}}
\newcommand{\JR}{J_{r}}

\newcommand{\h}[1]{\hat{#1}}
\newcommand{\dBLM}{\Delta B_{l}}
\newcommand{\dBMR}{\Delta B_{r}}

\newcommand{\vrf}{v_{\rm{rf}}}

\newcommand{\UDU}{\ket{\uparrow\downarrow\uparrow}}
\newcommand{\DUU}{\ket{\downarrow\uparrow\uparrow}}
\newcommand{\UUD}{\ket{\uparrow\uparrow\downarrow}}
\newcommand{\UUU}{\ket{\uparrow\uparrow\uparrow}}
\newcommand{\DDU}{\ket{\downarrow\downarrow\uparrow}}
\newcommand{\DUD}{\ket{\downarrow\uparrow\downarrow}}
\newcommand{\UDD}{\ket{\uparrow\downarrow\downarrow}}
\newcommand{\DDD}{\ket{\downarrow\downarrow\downarrow}}
\newcommand{\SL}{\ket{S_l}}
\newcommand{\SR}{\ket{S_r}}
\newcommand{\SM}{\ket{1}}

\newcommand{\TM}{\ket{0}}
\newcommand{\Q}{\ket{Q}}
\newcommand{\Qp}{\ket{Q_{+}}}

\newcommand{\eps}[2]{\varepsilon_{#2}^{\rm{#1}}} 
\newcommand{\tauJM}[1]{\tau_{\rm{#1}}}

\newcommand{\rhoZZ}{\ket{0}\!\!\bra{0}}

\newcommand{\w}[1]{\omega_{\mathrm{#1}}}
\begin{document}

\title{The Resonant Exchange Qubit}

\date{\today}

\author{J.~Medford$^{1}$}
\author{J.~Beil$^{2}$}
\author{J.~M.~Taylor$^{3}$}
\author{E.~I.~Rashba$^{1}$}
\author{H.~Lu$^4$}
\author{A.~C.~Gossard$^4$}
\author{C.~M.~Marcus$^{2}$}
\affiliation{$^1$Department of Physics, Harvard University, Cambridge, Massachusetts 02138, USA\\
$^2$Center for Quantum Devices, Niels Bohr Institute, University of Copenhagen, Universitetsparken 5, 2100 Copenhagen, Denmark\\
$^3$Joint Quantum Institute/NIST, College Park, MD, USA\\
$^4$Materials Department, University of California, Santa Barbara, California 93106, USA
}

\begin{abstract}
We introduce a solid-state qubit in which exchange interactions among confined electrons provide both the static longitudinal field and the oscillatory transverse field, allowing rapid and full qubit control via rf gate-voltage pulses.  We demonstrate two-axis control at a detuning sweet-spot, where leakage due to hyperfine coupling is suppressed by the large exchange gap. A $\pi/2$-gate time of 2.5~ns and a coherence time of 19 $\mu$s, using multi-pulse echo, are also demonstrated. Model calculations that include effects of hyperfine noise are in excellent quantitative agreement with experiment. 
\end{abstract}


\maketitle

As originally conceived, the two-level system that forms the basis of the semiconductor spin qubit is the electron spin itself, with pulsed exchange between two confined electrons forming a two-qubit gate~\cite{Loss_PRA98}. Generalizations to two-electron \cite{Petta_Science05, Taylor_NP05, Barthel_PRL09, Foletti_NP09, Shulman_Science12} and three-electron \cite{Divincenzo_Nature00,Gaudreau_PRL06,Gaudreau_APL09,Laird_PRB10,Gaudreau_NP11,Medford_preprint13,Braakman_arXiv13} qubits make use of multi-electron states as the quantum two-level system. These qubits offer ease of initialization, control, and readout, or speed of operation, in exchange for the complexity of controlling more than one electron per qubit. An attractive feature of the original single-spin proposal is that qubit rotations are implemented as Rabi processes, driven by a small resonant transverse field, rather than Larmor processes, which use pulsed Larmor precession around larger nonparallel fields. Rabi rotations allow narrow-band wiring away from dc, precession rates controlled by the amplitude of the oscillatory field, and straightforward two-axis control (needed for arbitrary transformations) implemented using the phase of the oscillatory field~\cite{Slichter,Vandersypen_RMP05}. 

In this Letter, we introduce a new quantum-dot-based qubit---the resonant exchange qubit---that captures the best features of previous incarnations, with qubit rotations via Rabi nutation using gate-controlled exchange both for the static longitudinal field and the oscillatory transverse field, as described in Ref.~\cite{Taylor_arXiv13}. The large exchange field suppresses leakage from the qubit space. However, because rotations are driven by a resonant transverse field, the large longitudinal field does not impose unrealistically fast evolution between qubit states. Moreover, the qubit is operated at a ``sweet spot'' of the exchange gap, making it insensitive to first order to electrical noise in the detuning parameter~\cite{Koch_PRA07,Schreier_PRB08,Paik_PRL11,Taylor_arXiv13}. 

\begin{figure}[b]
\includegraphics[width = 3.2 in]{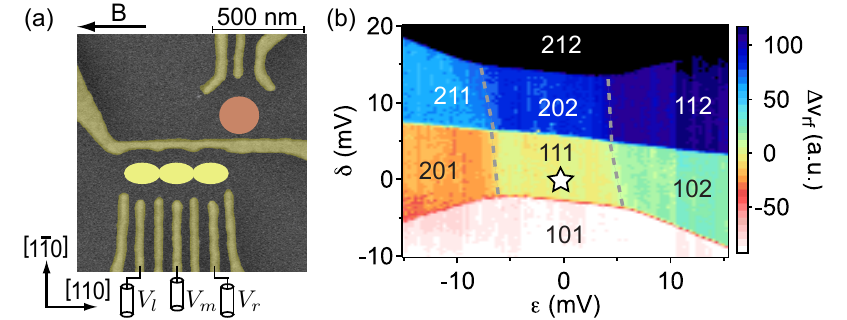}
\caption[Device and Charge Stability Diagram]{\label{FigDev}~(a)~False color micrograph of lithographically identical device with dot locations depicted; gates are marked in yellow. Gate voltages, $V_{l}$ and $V_{r}$, set the charge occupancy of left and right dot as well as the detuning, $\eps{}{}$ of the qubit. A neighboring sensor quantum dot is indicated with a larger circle. (b) Triple dot charge occupancy N$_{l}$ N$_{m}$ N$_{r}$ as a function of $\Vl$ and $\Vr$ in and near the 111 regime; $\eps{}{} = (\Vr -\Vr^{0})/2 - (\Vl-\Vl^0)/2$, $\delta =  (\Vr -\Vr^{0})/2 + (\Vl-\Vl^0)/2+\gamma (\Vm-\Vm^0)$. Measurements give $\gamma \sim 3$. The operating position is marked with a star, which is larger than the amplitude of voltage fluctuations used in rotations.}
\end{figure}

The resonant exchange qubit was realized in a triple quantum dot formed by surface gates 110 nm above a two-dimensional electron gas (density 2.6 $\times$ 10$^{15}$~m$^{-2}$, mobility 43~m$^2$/Vs) in a GaAs/Al$_{0.3}$Ga$_{0.7}$As heterostructure  [see Fig.~\ref{FigDev}(a)]. Gate voltages $\Vl$ and $\Vr$ controlled detuning, $\eps{}{} = (\Vr -\Vr^{0})/2 - (\Vl-\Vl^0)/2$, measured relative to the center of the 111 charge region, while $\Vm$ controlled the size of the 111 region (111 and other number triplets denote the charge occupancy of the triple dot)~\cite{epsFootnote}. An adjacent multi-electron quantum dot operated in Coulomb blockade regime served as a radio frequency (rf) charge sensor~\cite{Reilly_APL07,Barthel_PRB10}.

\begin{figure}
\includegraphics{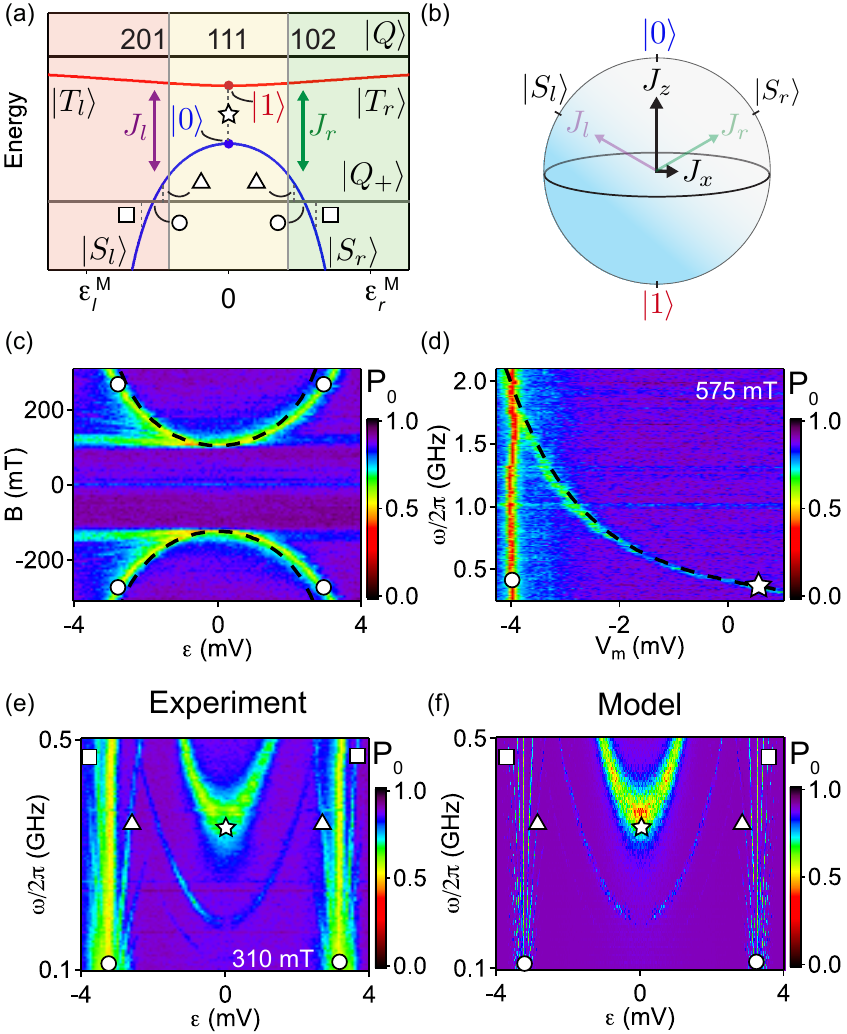} 
\caption[Qubit Spectroscopy]{
\label{FigSpec}~(a)~Energy level diagram for a constant $\delta$. Charge transitions are marked with circles (qubit-$\Qp$), triangles (qubit-$\Qp$, photon), squares (qubit-$\Qp$, photon), and a star ($\TM$-$\SM$) transitions. (b)~Schematic of the effects of $\JL$ and $\JR$ on the qubit Bloch sphere (c)~The qubit-$\Qp$ anti-crossing is mapped out in magnetic field and detuning without an excitation. Dashed line is a model of the exchange splitting for equal tunnel couplings. (d)~A sweep of the middle plunger gate at $\eps{}{}=0$ mV and fixed field of 575 mT, demonstrating control of the main qubit transition. The dashed curve is a model of $\JL$ + $\JR$ as a function of $\eps{}{0} = (\Vm-\Vm^0)/2$~\cite{tunnelFootnote}. (e)~At a fixed field of 310 mT, detuning and microwave burst frequency are swept to trace out the spectroscopy of the qubit. (f)~A model of qubit evolution in the presence of a microwave excitation and magnetic field gradients between dots in the longitudinal and transverse directions.
}
\end{figure}

Tunneling between adjacent quantum dots gives two exchange splittings, $\JL(\eps{}{})$, associated with the electron pair in the left and middle dots, and $\JR(\eps{}{})$, associated with the electron pair in middle and right dots. Away from zero detuning, defined as the center of 111, the qubit ground state, $\TM=\frac{1}{\sqrt{6}}(\UUD+\DUU-2\UDU)$, connects continuously to a singlet state of the left pair, $\SL=\frac{1}{\sqrt{2}}(\UDU-\DUU)$ in charge state 201, and to a singlet state of the right pair, $\SR=\frac{1}{\sqrt{2}}(\UUD-\UDU)$ in charge state 102. [see Fig.~2(a)]. The excited qubit state, $\SM=\frac{1}{\sqrt{2}}(\UUD-\DUU)$, maps into triplet states that, in contrast to the singlets, cannot tunnel into charge states 201 or 102.  This allows the qubit state to be detected with a charge sensor the distinguishes 201, 111, and 102. A third state, $\Qp = \UUU$, intersects the qubit ground state at two anti-crossings whose position depends on Zeeman splitting from an external magnetic field. By sweeping the magnetic field, the qubit ground-state energy can be measured as a function of detuning [Figs.~\ref{FigSpec}(a,c)].  The fourth state in Fig.~\ref{FigSpec}(a), $\Q=\frac{1}{\sqrt{3}}(\UUD+\UDU+\DUU)$, is separated from the qubit states by a sizable gap (half the separation between $\TM$ and $\SM$), suppressing leakage out of the qubit space. The gap to $\Q$ is deliberately kept large by setting tunneling rates, hence $\JL$ and $\JR$, to be large throughout the 111 charge region.

Qubit rotations are implemented by applying an oscillatory voltage to gate $\Vl$, which moves the operating point around $\eps{}{} = 0$, in turn creating an oscillatory transverse field $J_{x}$ [see Fig.~\ref{FigSpec}(b)]. When the oscillation frequency $\w{}$ matches the longitudinal exchange frequency, $J_{z}/\hbar$ [see Fig.~\ref{FigSpec}(b)], the qubit nutates between $\TM$ and $\SM$. Figure~\ref{FigSpec}(c) maps the positions of the $\Qp$ anti-crossings with the lower qubit branch as a function of field and detuning without applied microwaves, along with a model calculation of the exchange splittings $\JL$ and $\JR$. This spectroscopy is performed by preparing a $\SR$ state in 102, then pulsing into 111 for 300 ns before returning to 102 to project the resulting state back onto $\SR$.
 
 The data in Fig.~\ref{FigSpec}(d) shows two features, a vertical line corresponding to the crossing of $\Qp$ and the center of the lower qubit branch (circle), and a curved feature reflecting a driven oscillation between qubit states $\TM$ and $\SM$, marked with a star. The curved feature shows that the qubit splitting is controlled by gate voltage $\Vm$, here covering a range from 200 MHz to 2 GHz. Using fast gating, we have demonstrated control of this frequency on nanosecond time scales. The dashed line in Fig.~\ref{FigSpec}(d) is a model of $\w{}(V_m)$ that assumes a linear dependence of $\JL$ and $\JR$ on $\Vm$.
 
The resonant exchange qubit can be modeled by the Hamiltonian, 
\beq
	\mathcal{H}(\eps{}{}) = -J_z\sigma_z/2 - J_x\sigma_x/2,
\eeq
where $J_z = \frac{1}{2}(\JL(\eps{}{})+\JR(\eps{}{}))$ and $J_x= \frac{\sqrt{3}}{2}(\JR(\eps{}{})-\JL(\eps{}{}))$, where $\sigma_z = |0\rangle\langle 0| - |1\rangle\langle 1|$ and $\sigma_x = |0\rangle\langle 1| + |1\rangle\langle 0|$ are the Pauli operators of the qubit  [see Fig.~\ref{FigSpec}(b)]. Exchange fields $\JL(\eps{}{}) = -(\eps{}{}+\eps{}{0})/2 + \sqrt{t ^2+(\eps{}{}+\eps{}{0})^2/4}$ and $\JR(\eps{}{}) = (\eps{}{}-\eps{}{0})/2 + \sqrt{t ^2+(\eps{}{}-\eps{}{0})^2/4}$ are modeled in terms of the tunnel coupling, $t$, which is taken to be the same for both the 201-111 and 111-102 transitions, and $\pm\eps{}{0}$, the detunings of these charge transitions. At $\eps{}{}=0$, this gives $\mathrm{d}J_z/\mathrm{d}\eps{}{} = 0$ and $\mathrm{d}J_x/\mathrm{d}\eps{}{} = \frac{\sqrt{3}}{2}(1-\eps{}{0}/\sqrt{4 t^2+{\eps{}{0}}^2})$. For small detuning, $\eps{}{} \ll \eps{}{0}$, $J_z$ is unchanged to first order while $J_x\sim\eps{}{}$. This system is equivalent to a spin-1/2 in a large static field with a small transverse field. While $J_z$ is insensitive to detuning noise to first order, it is not insensitive to noise on gate $\Vm$ or other gates. However, other gates, including $\Vm$, do not need to operate at high frequency, and so can be heavily filtered.

\begin{figure}
\includegraphics[width = 2.5 in]{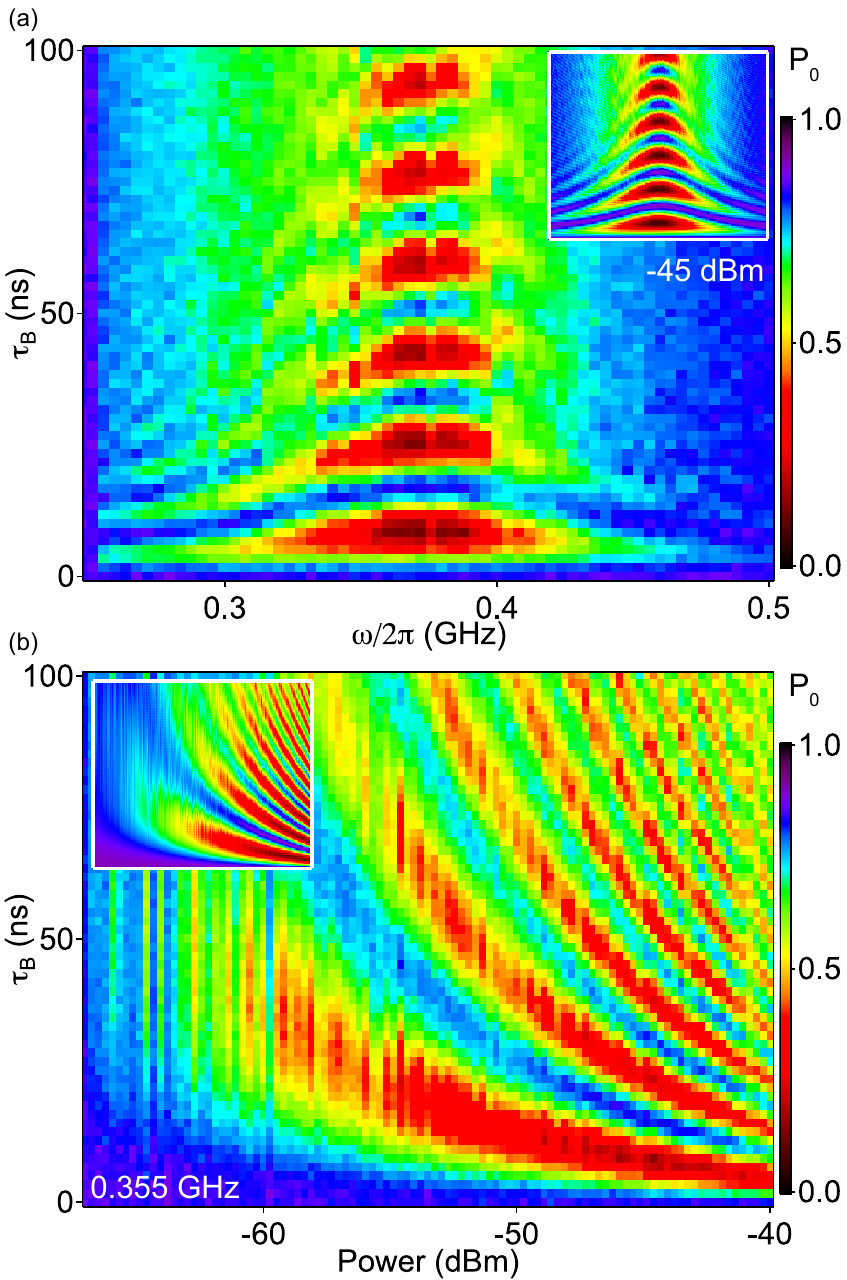}
\caption[Frequency and Amplitude Control]{
\label{FigRabi}~(a)~A Rabi nutation for a -45 dBm (0.45 mV) excitation on the left plunger gate with the detuning biased to the center of the transition [($\star$) in Fig.~\ref{FigSpec}(a,e)] for a time $\tauJM{B}$. (b)~A Rabi nutation with a 0.355 GHz excitation on the left plunger gate with the detuning biased to the center of the transition. (insets)~A model of this nutation using the exchange profile from Fig.~\ref{FigSpec}(f) and fluctuating longitudinal magnetic field gradients.
}
\end{figure}

In Fig.~\ref{FigRabi}, $\SR$ is prepared in 102 and adiabatically evolved to $\TM$ at $\eps{}{}=0$, taking care to move rapidly through the $\Qp$ anti-crossing. A microwave burst is then applied to $\Vl$ for a time $\tauJM{B}$ before returning adiabatically to 102 for measurement. The color plot shows the probability, $P_{0}$, of detecting the ground state through a charge measurement (see Sec.~I, Supplemental Material.) By sweeping frequency and power, we see patterns characteristic of Rabi nutations subject to low frequency noise in the splitting frequency, $\w{01}$ due to hyperfine gradients (see Sec.~VI, Supplemental Information). In the rotating frame, the amplitude of the oscillation gives the strength of the $\h{x}$ rotation, while the frequency detuning, $\delta = \w{}-\w{01}$, gives the strength of the $\h{z}$ rotation. As seen in Fig.~\ref{FigRabi}(b), as the power increases, effects of $\delta$ errors due to hyperfine gradients decrease.  At $\w{01}/2\pi = 0.355$ GHz, the nutation frequency scales with voltage as $\mathrm{d}\Omega_{R}/\mathrm{d}\Vl \sim 2\pi\times 70$ MHz/mV. This scaling increases with $\mathrm{d}J_x/\mathrm{d}\eps{}{}$, which grows as the 111 region is shrunk ($\eps{}{0}\rightarrow0$) to increase $\w{01}$. At $\w{01}/2\pi=1.98$ GHz, this scaling was measured to be $\sim 2\pi\times5$ GHz/mV, demonstrating a way to increase coupling to external voltages.

\begin{figure}
\includegraphics[width = 2.5 in]{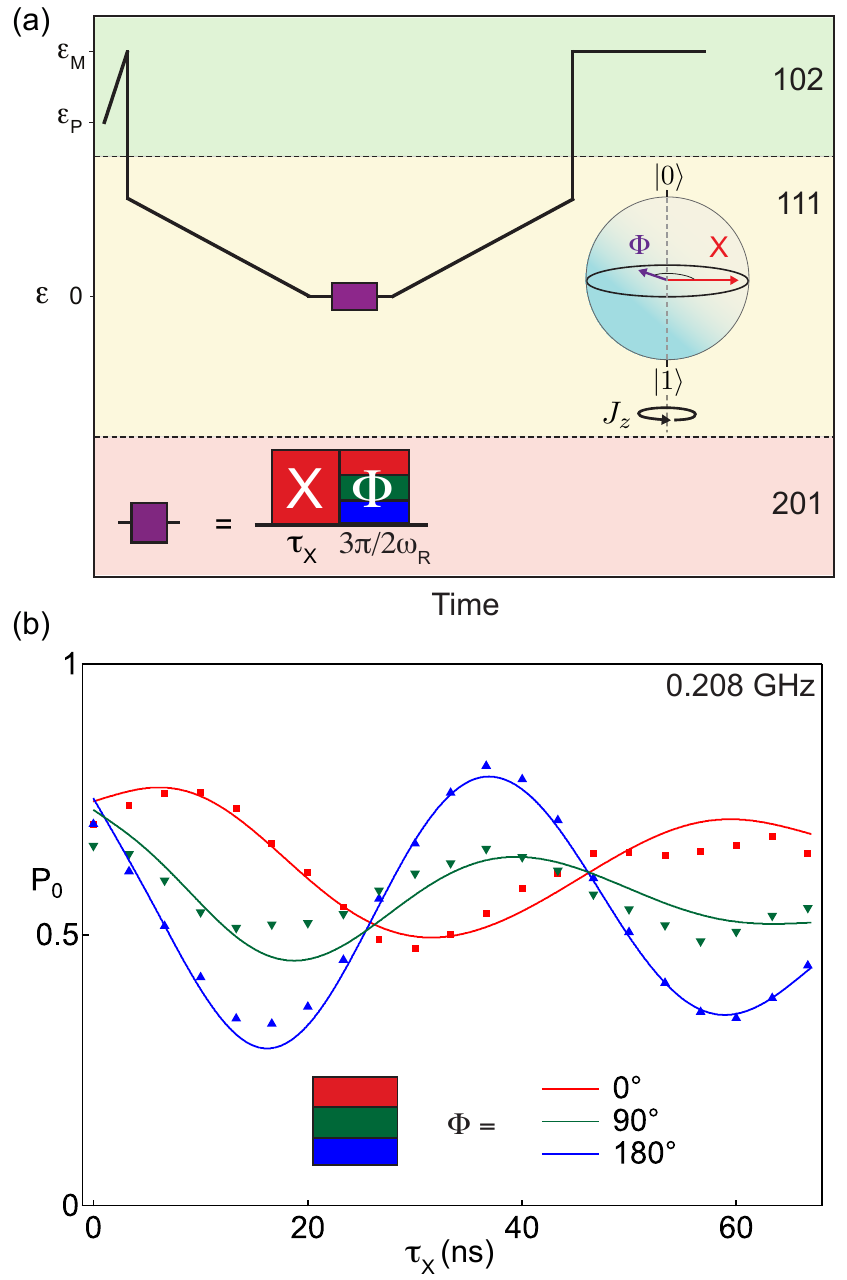}
\caption[Two-Axis Phase Control]{
\label{FigAngDep}~(a)~A schematic of the detuning during a two-pulse sequence, where the first pulse is an $X$ rotation and the second pulse is a rotation around an angle set by the relative phase of the carrier, $\Phi$, as depicted on the Bloch sphere. (b)~The qubit readout for a rotation about $X$, followed by a $\frac{3\pi}{2}$ rotation about an axis $\Phi$, for three different $\Phi$'s. The solid lines are fits to the model in Fig.~\ref{FigSpec}(c,d,f) and the insets of Fig.~\ref{FigRabi}.}
\end{figure}

On resonance in the rotating frame, the Hamilton takes the form $\mathcal{H}_{\mathrm{rf}} = \cos(\Phi)\sigma_x+\sin(\Phi)\sigma_y$, where $\Phi$ is the relative phase of the carrier wave with respect to the first pulse incident on the qubit.  Controlling phase relative to the initial pulse thus allows full two-axis qubit control. To test the qubit response, we prepare a $\TM$ and drive a rotation on resonance for a time $\tauJM{x}$, then apply a second pulse at relative phase $\Phi$ to drive a $3\pi/2$ rotation in a time $3\pi/2\w{R}$. Figure 4 shows data for $\Phi$ = $0^{\circ},$ $90^{\circ},$ and $180^{\circ}$, along with model curves using an optimized, though reasonable, value for hyperfine couplings as a fit parameter. 

\begin{figure}
\includegraphics[width = 2.9 in]{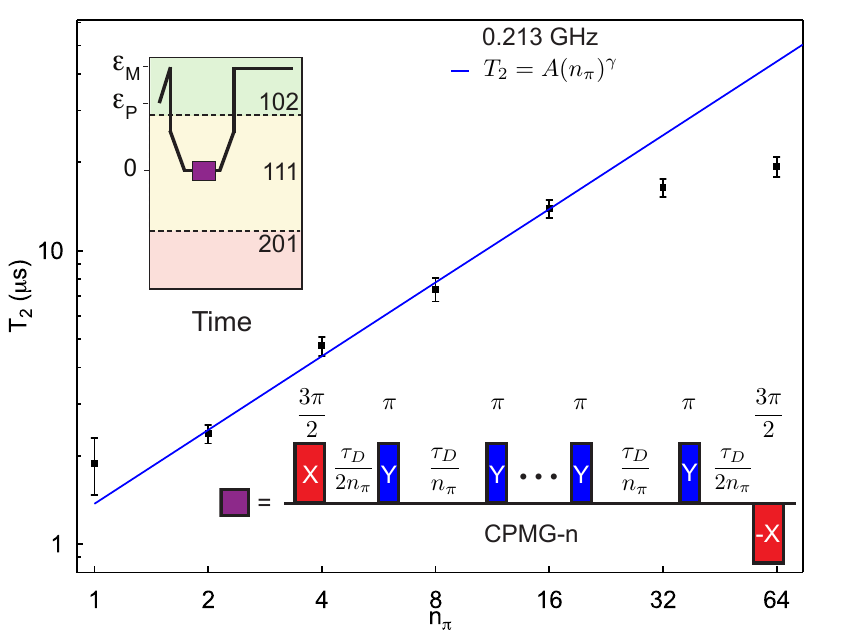}
\caption[Dynamical Decoupling and Noise Spectroscopy]{
\label{FigT2}~T$_2$  for various orders of CPMG-$n$, were each sequence contains $n$ $\pi$ rotations about $y$, as depicted in the lower inset. The upper inset depicts the detuning sequence for this experiment. We found that up to $n=16$, the even number of pulses was well described by T$_2 = A (n_{\pi})^\gamma$, where $\gamma = 0.84 \pm 0.05$. This translates to a power spectral density of $S(\w{})\sim \w{}^{-\beta}$, where $\beta = 5 \pm 1$.
 }
\end{figure}

Phase control was sufficient to implement a CPMG dynamical decoupling sequence, where $\pi$-pulses are applied along the $\h{y}$ axis in the rotating frame, partially decoupling rotation errors~\cite{Vandersypen_RMP05}. Figure~\ref{FigT2} shows resulting coherence time, T$_{2}$, for CPMG sequences up to 64 $\pi$-pulses, which gave T$_2 = 19 \pm 2\,\mu$s. Values for T$_{2}$ were extracted from Gaussian fits to $P_0(\tauJM{D})$, where $\tauJM{D}$ is the total dephasing time (see inset of Fig.~4).  Between 2 and 16 pulses, the scaling of coherence time with (even) pulse number, $n_{\pi}$, appears well described by the power-law, T$_2 = A (n_{\pi})^\gamma$, where $\gamma = 0.84\pm 0.05$.  Within a classical power-law noise model \cite{Cywinski_PRB08,Medford_PRL12} implies $S(\w{})\sim\w{}^{-\beta}$ with a $\beta= 5 \pm 1$. The inconsistency of this result with recent studies of electrical noise in the singlet-triplet qubit, where $\beta\sim 0.7$~\cite{Dial_arxiv12}, may reflect first-order insensitivity of the resonant exchange qubit to detuning noise. However, a detailed model for dynamical decoupling that distinguishes voltage noise from hyperfine noise has not been developed to date. Moreover, pulse sequences designed to decouple hyperfine noise for exchange-only qubits~\cite{Hickman_arXiv13} may also be adaptable to the resonant exchange qubit. 

For $n_{\pi}>16$, T$_2(n_{\pi})$ falls below the steep power-law, and appears to saturate around 20 $\mu$s. The measured T$_1$ for a splitting $\w{01}/2\pi = 0.33$ GHz was $\sim40 \,\mu$s, and decreased monotonically with increasing $\w{01}$, consistent with phonon-based relaxation, which suggests that T$_1$ was not limiting T$_2$ at $\w{01}/2\pi = 0.2$ GHz. Pulse errors are likely limiting T$_{2}$ in this measurement, though extending coherence much longer will require extending T$_1$.

In summary, we have introduced and demonstrated the operation of a new quantum-dot-based qubit that uses exchange for both the longitudinal and oscillatory transverse fields. A large exchange gap prevents state leakage, and the operating point is insensitive to first order to fluctuations in gate-controlled detuning. Two-axis control and a large ratio ($\sim 10^{4}$) of coherence time to gate operation time were demonstrated. Implementation of a two-qubit gate~\cite{Doherty_arXiv13,Taylor_arXiv13} is next experimental challenge. 

\textit{Acknowledgements}.---Research was supported in part by the Office of the Director of National Intelligence, Intelligence Advanced Research Projects Activity (IARPA), through the Army Research Office grant W911NF-12-1-0354. We also acknowledge support from the National Science Foundation Materials World Network Program, Harvard University, the Villum Foundation, and the Danish National Research Foundation, DARPA MTO and the NSF Physics Frontier Center at the JQI. We thank Maja Cassidy, Oliver Dial, David DiVincenzo, Andrew Doherty,  Mark Gyure, Bert Halperin, Ferdinand Kuemmeth, Thaddeus Ladd, and Arijeet Pal for useful discussions.

\setcounter{figure}{0}
\setcounter{equation}{0}
\renewcommand{\thefigure}{S\arabic{figure}}  
\renewcommand{\theequation}{S\arabic{equation}} 
\renewcommand{\figurename}{Figure} 
\renewcommand{\thefootnote}{S\arabic{reference}} 

\pagebreak
\onecolumngrid
\pagebreak
\section*{Supplementary Information for The Resonant Exchange Qubit}

\section{Measurement and Normalization}\label{SEOnorm}

A single normalization procedure was used for all data in the main paper to convert the measured reflectometry signals into output probabilities. It is similar to the normalization procedure described in  Ref.~\cite{Medford_preprint13Supp}, except that the normalization of the $T_1$ decay does not depend on measuring the overlap between $\SR$ and $\SL$. A measurement of this is impossible because the overlap $|\langle S_r\!\SL|^2$ requires diabatic passage through the center of the 111 region, and in our current setup the large gap in the center region forces our state to adiabatically follow the lower branch. Normalization of $T_1$ is done through a separate independent measurement of $T_1$ for each figure, as described below.

As in Ref.~\cite{Medford_preprint13Supp}, the measurements of a given parameter ($\eps{}{}$, $\w{}$, burst power, etc.) was repeated $2^{13}$ or $2^{14}$ times to obtain measurement statistics and then histogrammed, following the procedure in Ref.~\cite{Barthel_PRL09Supp}.  The resulting histogram is fit to a function of the form,
\begin{align}\label{eq:Hist}
 	n(\vrf) &= \frac{P}{\sqrt{2 \pi \sigma^2}}\,\exp\left[-\frac{(\vrf-\vrf^{102})^2}{2\sigma^2}\right]\nonumber\\
& + e^{-\tauJM{M}/T_1}\frac{(1-P)}{\sqrt{2 \pi \sigma^2}}\,\exp\left[-\frac{(\vrf-\vrf^{111})^2}{2\sigma^2}\right]\nonumber\\
& +\int_{\vrf^{102}}^{\vrf^{111}}\frac{d\rm{V}}{\sqrt{2 \pi \sigma^2}}\frac{\tauJM{M}}{T_1}\frac{(1-P)}{\Delta\vrf}\exp\left[-\frac{\tauJM{M}}{T_1}\frac{\rm{V}-\vrf^{102}}{\Delta\vrf}-\frac{(\vrf-\rm{V})^2}{2\sigma^2}\right],
\end{align}
where $n(\vrf)$ is the fraction of histogram events with outcomes $\vrf$ for a measurement in 102, $\vrf^{102}$ is the reflected voltage corresponding to double occupancy in the right dot, $\vrf^{111}$ is the reflected voltage corresponding to single charge occupancy in the all three dots,  $\Delta \vrf \equiv \vrf^{111}-\vrf^{102}$, $P$ is the fraction of 102 outcomes in the data set, $T_1$ is the relaxation time at at the measurement detuning $\eps{}{M}$, $\tauJM{M}$ is the measurement time, and $\sigma$ is the standard deviation of the histogram peaks due to noise in the rf equipment and shot noise intrinsic to the rf sensor dot. 

The extracted parameters $\vrf^{102}$ and $\vrf^{111}$ are then used to normalize the return probabilities $P$ as
\beq\label{eq:VrftoP}
	P  = \frac{\langle \vrf \rangle - \vrf^{111}}{\vrf^{102}-\vrf^{111}},
\eeq
where $\langle \vrf \rangle$ is the average voltage for a particular parameter ($\eps{}{}$, $\w{}$, burst power, etc.) over all repetitions of the measurement sequence. 

Equation \eqref{eq:VrftoP} converts $\vrf$ into a probability, but it does not account for relaxation during the measurement time $\tauJM{M}$, where a 111 state relaxes to a 102 state. As described in Ref.~\cite{Johnson_Nature05Supp}, $P$ is related to the actual probability $P_0$ through
\beq
	P_0 = \frac{1}{\tauJM{M}}\int^{\tauJM{M}}_{0} \mathrm{d}t\, P \exp\left(-\frac{t}{T_1}\right) =P \frac{T_1}{\tauJM{M}}\left[1-\exp\left(-\frac{\tauJM{M}}{T_1} \right) \right].
\eeq
By knowing $\tauJM{M}$ and $T_1$, we can correct for measurement relaxation. A measurement of the relaxation time at $\eps{}{M}$ is acquired for each section of data in the main paper, by fitting the average probability as a function of $\tauJM{M}$.

\section{Model of the exchange interactions}
We find that the detuning dependence of the exchange interactions $\JL$ and $\JR$ are well described by the model of 
\beq\label{eq:JL}
	\JL = -\frac{\alpha}{2} (\eps{}{}+\eps{}{0}) + \sqrt{\left\{t \exp\left[{-\left(\frac{\eps{}{}+\eps{}{0}}{W_t\eps{}{0}}\right)^2}\right]\right\}^2+\frac{\alpha^2}{4}(\eps{}{}+\eps{}{0})^2}
\eeq
\beq\label{eq:JR}
	\JR = \frac{\alpha}{2} (\eps{}{}-\eps{}{0}) + \sqrt{\left\{t \exp\left[{-\left(\frac{\eps{}{}-\eps{}{0}}{W_t\eps{}{0}}\right)^2}\right]\right\}^2+\frac{\alpha^2}{4}(\eps{}{}-\eps{}{0})^2},
\eeq
where $\alpha$ is the lever arm between $\eps{}{}$ and energy, $W_t$ is a phenomenological  suppression of the tunnel coupling with $\eps{}{}$, and $\pm \eps{}{0}$ is the detuning of the 111-102 and 111-201 charge transitions, or half the width of the 111 region. 
We can then write $J_z = \frac{1}{2}\left(\JL(\eps{}{})+\JR(\eps{}{}) \right)$ as
\begin{align}\label{eq:Jzbig}
	J_z &= -\frac{1}{2}\left(\alpha\eps{}{0}-\sqrt{\left\{t \exp\left[{-\left(\frac{\eps{}{}+\eps{}{0}}{W_t\eps{}{0}}\right)^2}\right]\right\}^2+\frac{\alpha^2}{4}(\eps{}{}+\eps{}{0})^2}\right.\nonumber \\
	&\qquad \left. {}- \sqrt{\left\{t \exp\left[{-\left(\frac{\eps{}{}-\eps{}{0}}{W_t\eps{}{0}}\right)^2}\right]\right\}^2+\frac{\alpha^2}{4}(\eps{}{}-\eps{}{0})^2}\right).
\end{align}
At $\eps{}{}=0$, eq~\eqref{eq:Jzbig} simplifies to
\beq\label{eq:JzofVm}
	J_z = -\alpha \frac{\Vm-\Vm^0}{4} + \sqrt{\left\{t \exp\left[{-\left(\frac{1}{W_t}\right)^2}\right]\right\}^2+\frac{\alpha^2}{16}(\Vm-\Vm^0)^2},
\eeq
where we have replaced $\eps{}{0}$ with $(\Vm-\Vm^0)/2$. Experimentally, we see that the width of the 111 region, $2\eps{}{0}$, is linear in $\Vm-\Vm^0$, with the same lever arm as the other gates. Equation~\eqref{eq:JzofVm} is used in Fig.~2(d) of the main text to map the resonance as a function of $\Vm$~\cite{tunnelConstantFootnote}. In Fig.~2(d), $t=16.9\,\mu$eV, $W_t$ was taken to be very large, such that the exponential was ignored, $\Vm^0$ was taken to be -4.05 mV on this plot~\cite{vMBackgroundFootnote}.

The transverse exchange, $J_x = \frac{\sqrt{3}}{2}\left(\JR(\eps{}{})-\JL(\eps{}{}) \right)$, can be written as
\begin{align}
	J_x &= \frac{\sqrt{3}}{2}\left(\alpha\eps{}{}-\sqrt{\left\{t \exp\left[{-\left(\frac{\eps{}{}+\eps{}{0}}{W_t\eps{}{0}}\right)^2}\right]\right\}^2+\frac{\alpha^2}{4}(\eps{}{}+\eps{}{0})^2}\right.\nonumber \\
	&\qquad \left. {}+ \sqrt{\left\{t \exp\left[{-\left(\frac{\eps{}{}-\eps{}{0}}{W_t\eps{}{0}}\right)^2}\right]\right\}^2+\frac{\alpha^2}{4}(\eps{}{}-\eps{}{0})^2}\right).
\end{align}

\section{Model used in Fig.~2(c) of the main text}
Reference~\cite{Laird_PRB10Supp} gives the separation between the lower branch of the qubit state, which they refer to as $\ket{\Delta^\prime}$, and the $\Q$ state as
\beq
	E_{\Delta^\prime Q} = -\frac{1}{2}\left(\JL+\JR+\sqrt{\JL^2+\JR^2-\JL\JR} \right),
\eeq
which is the difference between the lowest two qubit eigenvalues of equation~\eqref{eq:hJBig}. The separation between $\Q$ and $\Qp$, $E_{QQ+}$ is $g^*\mu B_{\mathrm{ext}}$. In Fig.~2(c) we plot the intersection of these two curves, $B_{\mathrm{ext}}(\eps{}{}) = B_0+ E_{\Delta^\prime Q}(\eps{}{})/\hbar g^*\mu$, where $B_0$ is an experimentally determined offset in the field due to remnant fields from ferromagnetic components in the cryostat. We find that an offset of $B_0 = -9.3$ mT and an effective g-factor of $g^*=-0.34$ describe our data well in the center of 111.
The tunnel coupling $t$ was 16.9 $\mu$eV, $\eps{}{0} = 3.7$ mV, $\alpha = 40\,\mu$eV/mV, $W_t = 3$.

\section{Model and Power Broadening in Fig.~2(d)}
\begin{figure}[ht!]
\centering
\includegraphics[width=5in]{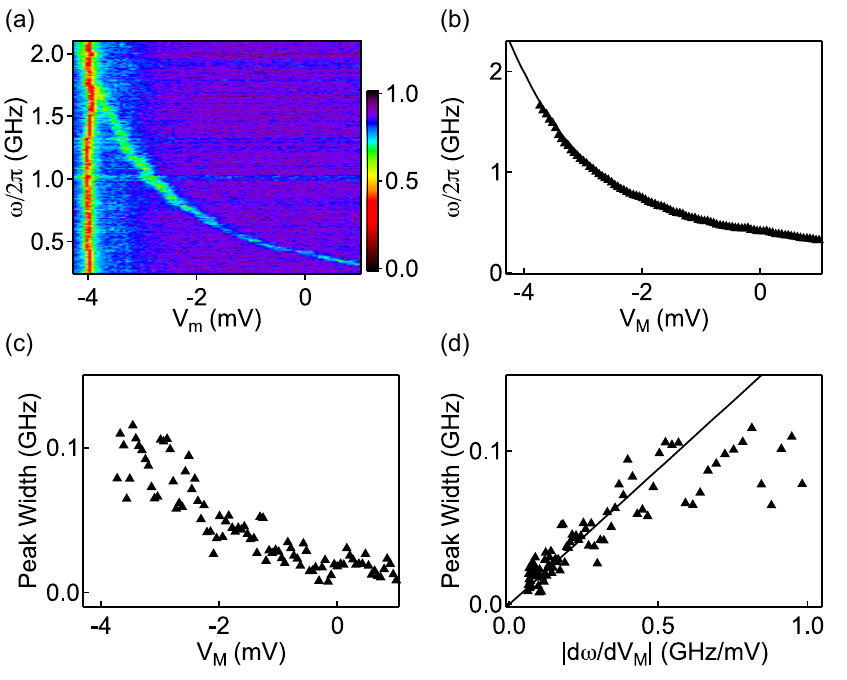} 
\caption[Power Broadening]{
\label{Fig:powerBroad}~(a)~ The data in Fig.~2(d) without the theory overlay. (b)~The resonance center value extracted from Gaussian fits, along with the plot of eq.~\eqref{eq:JzofVm}. (c)~The widths of the resonance extracted from the Gaussian fits. (d)~The widths of the resonance plotted against the analytic derivative of the qubit splitting $\w{} = J_z/\hbar $, where $J_z(\Vm)$ is given by eq.~\eqref{eq:JzofVm}. The analytic derivative was chosen to remove the noise from the numerical derivative of the data in panel (b).}
\end{figure}

The model in Fig.~2(d) is a plot of eq.~\eqref{eq:JzofVm}, where $\delta = (\Vl-\Vl^0)+(\Vr-\Vr^0)+\gamma(\Vm-Vm^0)$ was held constant, with an experimentally determined $\gamma=3$. In Fig.~\ref{Fig:powerBroad}(b), the resonance $\w{01}$ was extracted along with its width in frequency space by fitting it to a Gaussian at each value of $\Vm$. We find in Fig.~\ref{Fig:powerBroad}(d) that the resonance width, shown in Fig.~\ref{Fig:powerBroad}(c), is proportional to d$\w{01}$/d$\Vm$, which could suggest that the resonance widens with electrical noise. d$\w{01}$/d$\Vm$ is also proportional to d$J_x$/d$\eps{}{}$, which sets the strength of the Rabi oscillation. From this, we cannot determine whether the resonance is broadened due to fluctuations in $\Vm$, or due to power dependent broadening from an increased d$J_x$/d$\eps{}{}$.

\section{Model used in Fig.~2(f) of the main text}\label{sec:Manifold}
The model in Fig.~2(f) used seven of the eight spin states available to the three electron system to reproduce the spectroscopic data. \begin{align}
	\Qp \equiv \ket{Q_{+\frac{3}{2}}} &= \UUU\\
	\TM \equiv \ket{0_{+\frac{1}{2}}} &= \frac{1}{\sqrt{6}}\left(\UUD+\DUU-2\UDU\right)\\
	\SM \equiv \ket{1_{+\frac{1}{2}}} &= \frac{1}{\sqrt{2}}\left(\UUD-\DUU\right)\\
	\Q \equiv \ket{Q_{+\frac{1}{2}}} &= \frac{1}{\sqrt{3}}\left(\UUD+\DUU+\UDU\right)\\
	\ket{0_{-\frac{1}{2}}} &= \frac{1}{\sqrt{6}}\left(\DDU+\UDD-2\DUD\right)\\
	\ket{1_{-\frac{1}{2}}} &=\frac{1}{\sqrt{2}}\left(\DDU-\UDD\right)\\
	\ket{Q_{-\frac{1}{2}}} &= \frac{1}{\sqrt{3}}\left(\DDU+\UDD+\DUD\right)
\end{align}
\begin{figure}[ht!]
\centering
\includegraphics[width=3in]{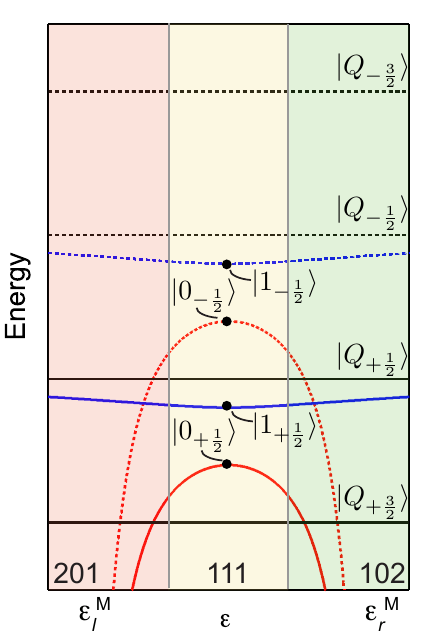} 
\caption{
\label{Fig:EnergySupp}~The full energy level spectrum of the three electron system. The dashed lines are states with opposite spin projection to those displayed in the main text. }
\end{figure}

\begin{figure}
\centering
\includegraphics[width=5.5in]{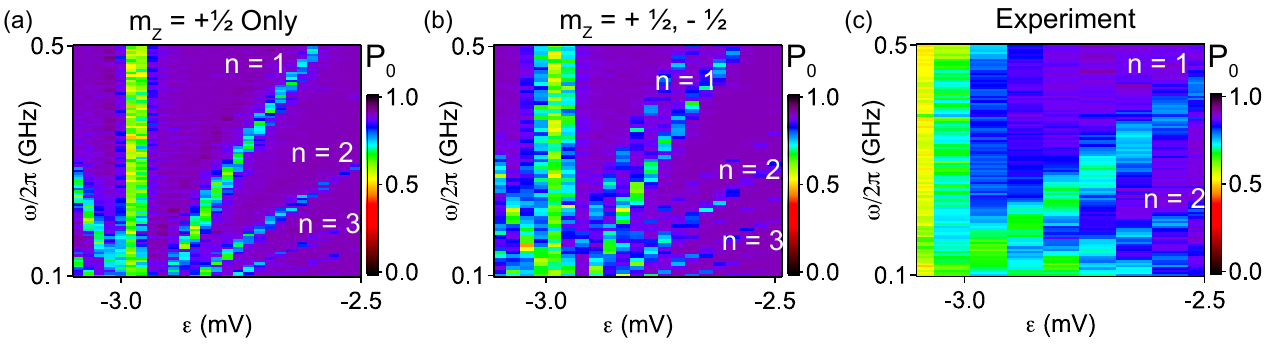} 
\caption[Spectroscopy with initialization in the upper Zeeman Manifold]{
\label{Fig:manifold}~A detailed view of the spectroscopy described in the main text in Figs.~2(e,f), near the anti-crossing between the lower qubit branch and $\ket{Q_{+\frac{3}{2}}}$ at $\eps{}{}\approx -3$ mV. (a)~A  simulation of the evolution of our three electron system near the $\ket{Q_{+\frac{3}{2}}}$ anti-crossing, where the initial state was contained in the $m_z = +1/2$ manifold $\left\{\ket{0_{+\frac{1}{2}}}, \ket{1_{+\frac{1}{2}}}, \ket{Q_{+\frac{1}{2}}}\right\}$. The slanting bright lines indicate transitions mediated by $n$-photon of frequency $\w{}$ between the lower qubit branch and $\ket{Q_{+\frac{3}{2}}}$, while the vertical bright line is the anti-crossing itself. Transitions to the right of the anti-crossing represent the stimulated emission of photons, while transitions to the left indicate the absorption of photons [see Fig.~2(a) of the main text and Fig.~\ref{Fig:EnergySupp}]. (b)~A simulation where the initial state was contained states from both the $m_z = +1/2$ and $m_z = -1/2$ manifolds $\left\{\ket{0_{+\frac{1}{2}}}, \ket{1_{+\frac{1}{2}}}, \ket{Q_{+\frac{1}{2}}},\ket{0_{-\frac{1}{2}}}, \ket{1_{-\frac{1}{2}}}, \ket{Q_{-\frac{1}{2}}}\right\}$. Here, the slanting bright lines are doubled, indicating an additional set of photon mediated transitions which are slightly offset in detuning~\cite{SamplingFootnote}. This implies that the presence of $m_z = -1/2$ states in the initial state should appear in the data as a double of the spectra around the $\ket{Q_{+\frac{3}{2}}}$ anti-crossing. (c)~A detailed look at the data corresponding to the same range of the simulation. The anti-crossing is shifted slightly from $\eps{}{}\approx -3$ mV due to small discrepancies between the exchange model and the data at large exchanges. The width of the photon lines in (c) is less than the spacing between the doubled lines in (b), leading us to conclude that $m_z = -1/2$ states were not loaded in significant amounts. A more detailed measurement which includes a spin selective readout of the third electron~\cite{Elzerman_Nature04Supp} would be necessary to confirm this result conclusively.}
\end{figure}

The states beyond the two levels of the qubit manifold are included to account for the possibility of nuclear mediated leakage from the qubit subspace $\left\{\ket{Q_{+\frac{3}{2}}}, \ket{Q_{+\frac{1}{2}}},\ket{Q_{-\frac{1}{2}}} \right\}$ as well as accidental initialization into the two states $\left\{\ket{0_{-\frac{1}{2}}},\ket{1_{-\frac{1}{2}}}\right\}$ that have the same total spin as the qubit states, $S_z = 1/2$, but opposite spin projection, $m_z = -1/2$, as shown in Fig.~\ref{Fig:EnergySupp}. In the regime that the device is operated in, the Zeeman splitting due to the applied magnetic field is less then the electron temperature, preventing us from preferentially loading $\ket{0_{+\frac{1}{2}}}$ and $\ket{1_{+\frac{1}{2}}}$ instead of $\ket{0_{-\frac{1}{2}}}$ and $\ket{1_{-\frac{1}{2}}}$. The simulations presented in Fig.~\ref{Fig:manifold} give some indication that by avoiding replenishing our electrons from the leads we can in fact prepare $\ket{0_{+\frac{1}{2}}}$ and $\ket{1_{+\frac{1}{2}}}$, which allows us to ignore the higher energy states, $\left\{\ket{0_{-\frac{1}{2}}},\ket{1_{-\frac{1}{2}}},\ket{Q_{-\frac{1}{2}}} \right\}$ in subsequent simulations.

The eighth level, $\ket{Q_{-\frac{3}{2}}}=  \DDD$, is separated from all other levels by the external magnetic field as shown in Fig.~\ref{Fig:EnergySupp}, and is therefore ignored in order to speed up computation.
The simulation described the qubit evolution in the presence of exchange interactions and Zeeman energy from longitudinal and transverse nuclei, which we account for with the following Hamiltonians written in the basis of \\
$\left\{\ket{Q_{+\frac{3}{2}}},\ket{0_{+\frac{1}{2}}},\ket{1_{+\frac{1}{2}}}, \ket{Q_{+\frac{1}{2}}},\ket{0_{-\frac{1}{2}}},\ket{1_{-\frac{1}{2}}},\ket{Q_{-\frac{1}{2}}}\right\}$ as:
\beq\label{eq:hJBig}
\h{H}_{J} =
\bep
0 & 0 & 0 & 0 & 0 & 0 & 0  \\
 0 & -\frac{3(\JL+\JR)}{4}  & \frac{\sqrt{3}(\JL-\JR)}{4} &0 & 0 & 0 & 0\\
 0 & \frac{\sqrt{3}(\JL-\JR)}{4} & -\frac{\JL+\JR}{4} & 0& 0 & 0 & 0 \\
 0 & 0 & 0 &0 & 0 & 0 & 0 \\
 0 & 0 & 0 & 0 &-\frac{3(\JL+\JR)}{4} & \ \frac{\sqrt{3}(\JL-\JR)}{4} &0 \\
 0 & 0 & 0 & 0 & \frac{\sqrt{3}(\JL-\JR)}{4} & -\frac{\JL+\JR}{4} & 0\\
 0 & 0 & 0 & 0 & 0 &0 &0   \\
\eep,
\eeq
\beq\label{eq:hBZ}
\h{H}_{Bz} =g^*\mu
\left(
\begin{smallmatrix}
 \frac{Bz_1+Bz_2+Bz_3}{2} & 0 & 0 & 0 & 0 & 0 & 0  \\
 0 & \frac{2 Bz_1-Bz_2+2 Bz_3}{6}  & \frac{Bz_1-Bz_3}{2 \sqrt{3}} & -\frac{Bz_1-2 Bz_2+Bz_3}{3 \sqrt{2}} & 0 & 0 & 0\\
 0 & \frac{Bz_1-Bz_3}{2 \sqrt{3}} & \frac{Bz_2}{2} & \frac{Bz_1-Bz_3}{\sqrt{6}} & 0 & 0 & 0 \\
 0 & -\frac{Bz_1-2 Bz_2+Bz_3}{3 \sqrt{2}} & \frac{Bz_1-Bz_3}{\sqrt{6}} & \frac{Bz_1+Bz_2+Bz_3}{6} & 0 & 0 & 0 \\
 0 & 0 & 0 & 0 & \frac{-2 Bz_1+Bz_2-2 Bz_3}{6}  & \frac{-Bz_1+Bz_3}{2 \sqrt{3}} & \frac{Bz_1-2 Bz_2+Bz_3}{3 \sqrt{2}} \\
 0 & 0 & 0 & 0 & \frac{-Bz_1+Bz_3}{2 \sqrt{3}} & -\frac{Bz_2}{2} & \frac{-Bz_1+Bz_3}{\sqrt{6}}\\
 0 & 0 & 0 & 0 & \frac{Bz_1-2 Bz_2+Bz_3}{3 \sqrt{2}} & \frac{-Bz_1+Bz_3}{\sqrt{6}} & \frac{-Bz_1-Bz_2-Bz_3}{6}   \\
\end{smallmatrix}
\right),
\eeq
\beq\label{eq:hBX}
\h{H}_{Bx} = g^*\mu
\left(
\begin{smallmatrix}
 0 & \frac{Bx_1-2 Bx_2+Bx_3}{2 \sqrt{6}} & \frac{-Bx_1+Bx_3}{2 \sqrt{2}} & \frac{Bx_1+Bx_2+Bx_3}{2 \sqrt{3}} & 0 & 0 & 0 \\
 \frac{Bx_1-2 Bx_2+Bx_3}{2 \sqrt{6}} & 0 & 0 & 0 & \frac{-2 Bx_1+Bx_2-2 Bx_3}{6}  & \frac{-Bx_1+Bx_3}{2 \sqrt{3}} & -\frac{Bx_1-2 Bx_2+Bx_3}{6 \sqrt{2}}  \\
 \frac{-Bx_1+Bx_3}{2 \sqrt{2}} & 0 & 0 & 0 & \frac{-Bx_1+Bx_3}{2 \sqrt{3}} & -\frac{Bx_2}{2} & \frac{Bx_1-Bx_3}{2 \sqrt{6}}\\
 \frac{Bx_1+Bx_2+Bx_3}{2 \sqrt{3}} & 0 & 0 & 0 & -\frac{Bx_1-2 Bx_2+Bx_3}{6 \sqrt{2}} & \frac{Bx_1-Bx_3}{2 \sqrt{6}} & \frac{Bx_1+Bx_2+Bx_3}{3}  \\
 0 & \frac{-2 Bx_1+Bx_2-2 Bx_3}{6}  & \frac{-Bx_1+Bx_3}{2 \sqrt{3}} & -\frac{Bx_1-2 Bx_2+Bx_3}{6 \sqrt{2}} & 0 & 0 & 0\\
 0 & \frac{-Bx_1+Bx_3}{2 \sqrt{3}} & -\frac{Bx_2}{2} & \frac{Bx_1-Bx_3}{2 \sqrt{6}} & 0 & 0 & 0 \\
 0 & -\frac{Bx_1-2 Bx_2+Bx_3}{6 \sqrt{2}} & \frac{Bx_1-Bx_3}{2 \sqrt{6}} & \frac{Bx_1+Bx_2+Bx_3}{3} & 0 & 0 & 0 \\
\end{smallmatrix}
\right),
\eeq
and $\h{H}_{By} = g^*\mu$
\beq\label{eq:hBY}
\left(
\begin{smallmatrix}
 0 & -\frac{i (By_1-2 By_2+By_3)}{2 \sqrt{6}} & \frac{i (By_1-By_3)}{2 \sqrt{2}} & -\frac{i (By_1+By_2+By_3)}{2 \sqrt{3}} & 0 & 0 & 0  \\
 \frac{i (By_1-2 By_2+By_3)}{2 \sqrt{6}} & 0 & 0 & 0 & \frac{ i (2 By_1-By_2+2 By_3)}{6} & \frac{i (By_1-By_3)}{2 \sqrt{3}} & \frac{i (By_1-2 By_2+By_3)}{6 \sqrt{2}}  \\
 -\frac{i (By_1-By_3)}{2 \sqrt{2}} & 0 & 0 & 0 & \frac{i (By_1-By_3)}{2 \sqrt{3}} & \frac{i By_2}{2} & -\frac{i (By_1-By_3)}{2 \sqrt{6}}  \\
 \frac{i (By_1+By_2+By_3)}{2 \sqrt{3}} & 0 & 0 & 0 & \frac{i (By_1-2 By_2+By_3)}{6 \sqrt{2}} & -\frac{i (By_1-By_3)}{2 \sqrt{6}} & -\frac{i (By_1+By_2+By_3)}{3}   \\
 0 & -\frac{i (2 By_1-By_2+2 By_3)}{6}  & -\frac{i (By_1-By_3)}{2 \sqrt{3}} & -\frac{i (By_1-2 By_2+By_3)}{6 \sqrt{2}} & 0 & 0 & 0  \\
 0 & -\frac{i (By_1-By_3)}{2 \sqrt{3}} & -\frac{i By_2}{2} & \frac{i (By_1-By_3)}{2 \sqrt{6}} & 0 & 0 & 0 \\
 0 & -\frac{i (By_1-2 By_2+By_3)}{6 \sqrt{2}} & \frac{i (By_1-By_3)}{2 \sqrt{6}} &  \frac{i(By_1+By_2+By_3)}{3}  & 0 & 0 & 0  \\
\end{smallmatrix}
\right).
\eeq
Here, $g^* \approx -0.34$, as determined from Figs.~2(c,e). The magnetic field terms $Bx_i$, $By_i$, and $Bz_i$ in equations \eqref{eq:hBX}, \eqref{eq:hBY}, and \eqref{eq:hBZ} are the magnetic fields along $\hat{x}$, $\hat{y}$, $\hat{z}$ respectively in dot $i$, where $i=1$ corresponds to the left. The exchange terms $\JL$ and $\JR$ in eq.~\eqref{eq:hJBig} are the $\eps{}{}$-dependent terms from eqs.~\eqref{eq:JL} and \eqref{eq:JR}.

The model was created in the following way. At a given $\eps{}{}$ and $\w{}$, nine random variables were drawn from a normal distribution to take the nine nuclear field components, $B_x$, $B_y$, and $B_z$ in each of the three dots. From there, eigenstates of the full Hamiltonian, $\h{H} = \h{H}_J+\h{H}_{Bx}+\h{H}_{By}+\h{H}_{Bz}$, were calculated, and an initial state density matrix was chosen as a mixture of $90\%$ of the eigenstate with the largest overlap with $\ket{0_{+\frac{1}{2}}}$, and $5\%$ of the eigenstates with the largest overlaps with $\ket{1_{+\frac{1}{2}}}$ and $\ket{Q_{+\frac{1}{2}}}$ to account for loading infidelity. The initial state was then time evolved according to the Liouville-von Neumann equation, 
\beq\label{eq:LvN}
i \hbar\frac{\mathrm{d}\rho}{\mathrm{d}t} =  [\h{H},\rho],
\eeq
 for 300 ns in the presence of an oscillatory $\eps{}{}$. The final density matrix was then transformed into eigenstates of only the exchange interactions and the external magnetic field, and the population of the lower qubit eigenstate was recorded~\cite{SimFootnote1}. This process is then repeated 25 times with new random values for all of the nuclear fields, and the average return probability is recorded in the model~\cite{averagingFootnote}. 
 
In the model, the amplitude was a 0.225 mV oscillation in detuning, equivalent to a 0.45 mV oscillation in $\Vl$, or $-51$ dBm. The standard deviation of nuclear gradients was 3.9 mT with a $g^*=-0.34$. The tunnel coupling $t$ was 16.9 $\mu$eV, $\eps{}{0} = 3.7$ mV, $\alpha = 40\,\mu$eV/mV, $W_t = 3$.

The simulation used seven levels to enable us to check whether we loaded in states from the $m_z=-1/2$ manifold $\left\{\ket{0_{-\frac{1}{2}}},\ket{1_{-\frac{1}{2}}},\ket{Q_{-\frac{1}{2}}}\right\}$. Figure~\ref{Fig:manifold}(a) shows a detailed view of the region near the $\ket{Q_{+\frac{3}{2}}}$ anti-crossing from the simulation Fig.~2(f) in the main text. Panel (b) shows that same region with a mixture of both $m_z=+1/2$ and $m_z=-1/2$ states loaded into the initial state, approximating the density matrix in the case that we load into the other Zeeman manifold approximately half the time. The presence of these extra states causes a doubling of the transition lines near this anti-crossing, a phenomenon that we do not see reproduced in the data, shown in Fig.~\ref{Fig:manifold}(c). From this we determine that we are only loading in the $m_z=+1/2$ manifold. This is consistent with our initialization procedure which maintains isolation from the higher temperature leads.

\section{Model used in Fig.~3 insets of the main text}
Having shown in the previous section that we are only in the space of $\ket{0_{+\frac{1}{2}}}$, $\ket{1_{+\frac{1}{2}}}$, and $\ket{Q_{+\frac{1}{2}}}$ states, we now restrict ourselves to only the appropriate $3\times3$ subregions of Hamiltonians~\eqref{eq:hJBig} and \eqref{eq:hBZ}, which is valid when we are far away from the $\Qp$ anti-crossings as we are in Fig.~3. We express these as
\beq\label{eq:hJSmall}
\h{H}_{J} =
\bep
 -\frac{3(\JL+\JR)}{4}  & \frac{\sqrt{3}(\JL-\JR)}{4} &0 \\
 \frac{\sqrt{3}(\JL-\JR)}{4} & -\frac{\JL+\JR}{4} & 0 \\
  0 & 0 &0  \\
\eep,
\eeq
\beq\label{eq:hBZSmall}
\h{H}_{Bz} =g^*\mu
\bep
  \frac{2 Bz_1-Bz_2+2 Bz_3}{6}  & \frac{Bz_1-Bz_3}{2 \sqrt{3}} & -\frac{Bz_1-2 Bz_2+Bz_3}{3 \sqrt{2}} \\
 \frac{Bz_1-Bz_3}{2 \sqrt{3}} & \frac{Bz_2}{2} & \frac{Bz_1-Bz_3}{\sqrt{6}}\\
  -\frac{Bz_1-2 Bz_2+Bz_3}{3 \sqrt{2}} & \frac{Bz_1-Bz_3}{\sqrt{6}} & \frac{Bz_1+Bz_2+Bz_3}{6}  \\
\eep,
\eeq
with $\h{H}_{Bx} = 0$, and $\h{H}_{By} = 0$. 
We also revert to the simpler notation of $\TM$, $\SM$, and $\Q$. For the insets, we create an initialization state that is 90\% $\TM$, and 10\% $\SM$ and $\Q$, and time evolve it using equation \eqref{eq:LvN} for 100 ns in the presence of an oscillatory $\eps{}{}$ and static longitudinal nuclear field gradients $\dBLM$ and $\dBMR$. The $\rhoZZ$ term is extracted at multiple times during the evolution, and recorded. This process is repeated with a new set of $\dBLM$ and $\dBMR$ drawn from a random normal distribution, and averaged with the previous set. Unlike the simulation in Fig.~2(f), the nuclei are presumed to be static over the course of a single column, which provides the flickering effect seen in the model and the data. The data was acquired more rapidly in this set, justifying the modification to the model.

In the model in Fig.~3(a), the amplitude was a 0.45 mV oscillation in detuning, equivalent to a 0.89 mV oscillation in $\Vl$, or $-45$ dBm. The standard deviation of nuclear gradients was 3.9 mT with a $g^*=-0.34$. The tunnel coupling $t$ was 16.9 $\mu$eV, $\eps{}{0} = 3.7$ mV, $\alpha = 40\,\mu$eV/mV, $W_t = 3$.

\section{Model used in Fig.~4 of the main text}
The theory curves in Fig.~4 are based on a similar model to the insets in Fig.~3, with the complication that we perform a second time evolution which takes the final state of the first evolution as its input. In the second time evolution, the $\eps{}{}$ oscillation at a phase $\Phi$ with respect to the oscillation in the first time evolution. The preparation was 90\% $\TM$ and 10\% $\SM$ and $\Q$, and both longitudinal as well as transverse nuclear fluctuations were incorporated in a quasi-static manner, as they were in the inset of Fig.~3.

In the model curves in Fig.~4(b), the amplitude was a 0.30 mV oscillation in detuning, equivalent to a 0.60 mV oscillation in $\Vl$, or $\sim-55$ dBm. The standard deviation of nuclear gradients was 5.2 mT with a $g^*=-0.34$. The tunnel coupling $t$ was 12.4 $\mu$eV, $\eps{}{0} = 3.7$ mV, $\alpha = 40\,\mu$eV/mV, $W_t = 3$. The somewhat larger standard deviation of nuclear gradients in this data compared to the data in Fig.~3 is not well understood, but since the nuclear field fluctuates slowly on the order of minutes, we expect to observe a range of standard deviations among individual data sets.


\begin{thebibliography}{27}
\bibitem[{tun()}]{tunnelFootnote}
\bibinfo{note}{The theory curve is a plot of $\w{01}
  =\left(-\beta(\Vm-\Vm^0)+\sqrt{4t^2+\beta^2(\Vm-\Vm^0)^2}\right)/4\hbar$,
  where $\beta$ = 20 $\mu$eV/mV, $t$ = 16.9 $\mu$eV, and $\Vm^0$ = -4.05 mV. A
  constant tunnel coupling was used here rather than a Gaussian dependent
  tunnel coupling, because $\Vl$ and $\Vr$ were constant.}
\end{thebibliography}
\begin{thebibliography}{10}
\bibitem[{tun()}]{tunnelConstantFootnote}
\bibinfo{note}{In Fig.~2(d), the tunnel coupling was simplified back to just
  $t$, with no exponential dependence. The physical reason for this is not well
  understood, but the model clearly fits the data if this is true, and
  drastically disagrees if the exponential dependence is included.}
\bibitem[{vMB()}]{vMBackgroundFootnote}
\bibinfo{note}{An overall background to $\Vm$ of $-452$ mV was already removed,
  leaving just the amount that was pulsed using the AWG.}
\bibitem[{Sim()}]{SimFootnote1}
\bibinfo{note}{The eigenstates of the Hamiltonian with the nuclei were used for
  the initialization because the slow ramp in to this state allows the system
  to adiabatically enter the anti-crossing with $\ket{Q_{\frac{3}{2}}}_+$. The
  pulse to measurement position is a much more rapid, meaning that the qubit
  left that region diabatically. The difference in eigenstates allows us to
  capture this effect reasonably well without considering the full time
  dynamics of the problem.}
\bibitem[{ave()}]{averagingFootnote}
\bibinfo{note}{In order to speed up computation, the repetitions were truncated
  at 5 samples if the average return probability was $>86\%$. This saw a
  drastic improvement in speed with little qualitative change to the plots.}
\bibitem[{Sam()}]{SamplingFootnote}
\bibinfo{note}{Panel (b) was sampled more sparsely because it was a test run of
  the larger initialization space. Since it did not accurately represent the
  data, it did not receive a follow up calculation at higher sampling, as panel
  (a) did.}
\end{thebibliography}
\end{document}